\documentclass[aps,prb,twocolumn,amsmath,amssymb, superscriptaddress,tightenlines]{revtex4}
\usepackage{graphicx}
\usepackage{epsfig}
\usepackage{dcolumn}
\usepackage{bm}
\usepackage{amssymb}
\usepackage{bm}
\usepackage{amsmath, bm}
\usepackage{upgreek}
\usepackage[colorlinks,citecolor=blue,linkcolor=blue,hyperindex]{hyperref}

\begin{document}

\title{Robustness of quantum Hall interferometry}

\author{D. E. Feldman}
\affiliation{Department of Physics, Brown University, Providence, Rhode Island 02912, USA}
\affiliation{Brown Theoretical Physics Center, Brown University, Providence, Rhode Island 02912, USA}
 \author {Bertrand I. Halperin}
\affiliation{Department of Physics, Harvard University, Cambridge, Massachusetts 02138, USA}  
\date{\today}

\begin{abstract}
{\color{black}Fabry-P\'{e}rot  interferometry has emerged as a tool to probe anyon statistics in the quantum Hall effect. The interference phase is interpreted as a combination of a quantized statistical phase and an Aharonov-Bohm phase, proportional to the device area and the charge of the anyons propagating along the device edge. This interpretation faces two challenges.
First, the edge states have a finite width and hence the device area is ill-defined. Second, multiple localized anyons may be present in states that overlap with the edge, and it may not be clear whether a second anyon traveling along the edge will go inside or outside {\color{black} the region with a} localized anyon and therefore whether or not it should pick up a statistical phase.  
 We show how one may  overcome both challenges. In a case where only one chiral edge mode passes through the constrictions defining the interferometer, as when electrons in  a constriction are in  a Laughlin state with $\nu=1/(2n+1)$ or the integer state at $\nu=1$, we show that the interference phase can be directly related to the total electron charge contained in the interferometer. This holds for arbitrary electron-electron interactions and holds even if the bulk of the interferometer has a higher electron density than the region of the constrictions.  
In contrast to the device area or to the number of anyons inside a propagating edge channel,  the total  charge is well-defined.  We examine, at the microscopic level, how the relation between charge and phase is maintained when  there is a soft confining potential and disorder near the edge of the interferometer, and we discuss briefly the complications that can occur when multiple chiral modes can pass through the constriction.}
\end{abstract}

\maketitle

\section{Introduction}

Arguably, one deals with quantum interference every time the Schr{\"o}dinger equation is being solved. However, thinking about the solution in such a language is not always illuminating. This language is most natural when a small number of well-defined paths with the same starting and ending points are available as is the case 
in quantum Hall interferometry \cite{review-FH}. Fig. 1 illustrates a typical setup. Charge can travel only counter-clockwise along the chiral edges of the sample. Tunneling is possible at two constrictions. Hence, two interfering paths connect the source and the detector. 

This setup allows probing fractional charge and statistics through the interference phase \cite{review-FH}. {\color{black}It is generally understood that} the phase contains three components: a non-universal phase from the tunneling amplitudes in the constrictions, the Aharonov-Bohm phase, proportional to the tunneling charge and the magnetic flux through the device, and a statistical phase that depends on the number of the localized quasiparticles in the interferometer and their mutual statistics with the tunneling particle. The three phases cannot be disentangled at a fixed magnetic field and fixed gate voltages, but respond in different ways to a change of the magnetic field. The non-universal contribution depends only weakly on the field, the Aharonov-Bohm phase changes continuously, and the statistical phase jumps every time a new quasiparticle enters the device \cite{int1,int2}. This idea has been recently used to directly observe \cite{manfra20} the statistics of anyons at the filling factor $1/3$. Several other experiments \cite{w52-1,w52-2,w52-3} have been interpreted in the same physical picture \cite{eo1,eo2}. It was also proposed to extend interferometry to heat currents \cite{added-WMF,added}.

\begin{figure}[!htb]
\bigskip
\centering\scalebox{0.35}[0.35]
{\includegraphics{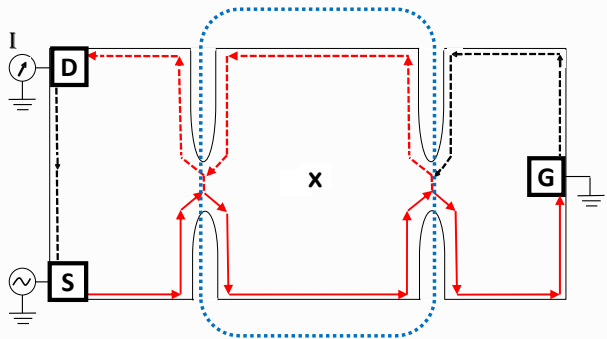}}
\caption{Schematic of a quantum Hall  interferometer. Charge travels counterclockwise along chiral edges and can tunnel between the opposite edges at two constrictions. Two paths, indicated by the dashed red line, allow charges from the  source S to reach the detector contact D, after tunneling across one of the constrictions.
 In the absence of tunneling,  charges from S  would follow  the red solid  path to the ground contact G.  Black dashed lines are at ground potential and carry no current.   A cross shows an anyon localized in the device.  The dotted curve is a contour which will be used to define the charge on the interferometer in Section III.     }
\label{fig1:inter}
\end{figure}

Yet, the above physical picture faces two challenges. First, the edge has a finite width \cite{CSG,GH} and it is unclear why the Aharonov-Bohm phase is well defined, even for an integer quantum Hall state. Second, the statistical phase is defined for anyons, moving in the gapped bulk of a quantum Hall liquid around a remote anyon \cite{review-FH}. In interferometry, the interfering changes move along gapless edges.  In GaAs, those edges run in the region of a lower filling factor than in the bulk \cite{CSG,GH}. This should be understood as containing a finite concentration of quasiholes. Hence, {\color{black} a fluctuating number of  anyons may}  be  present near the edges. It is not obvious that the statistical phase is robust in such a situation.

{\color {black} 
The purpose of this paper is to explore these questions.  We concentrate on the simplest cases, such as the integer state at $\nu=1$ and the Laughlin states with $\nu = 1/(2n+1)$, which have only a single chiral edge state in a system with sharp edges and no disorder.  We find that the  interference phase in these cases should  be remarkably robust in the presence of disorder and a soft confining potential at the sample edges, in part because the interference phase can be directly related to the total electric charge in the interferometer  region.   

In Section II, below,  we examine the edge of  semi-infinite quantum Hall system in a situation where there is disorder and soft edge confinement.  We discuss what would be expected in a Hartree-Fock approximation, and then turn to a bosonized description of the edge mode.  

In Section III, we apply these results to the geometry of a Fabry-P\'erot interferometer.
We show that the  interference phase measured  in an interferometer experiment at low current and low temperature is directly related to the phase accumulation around the interferometer edge in the electronic ground state.  We also establish the result that as long as there is only a single chiral mode that can pass through the constrictions defining the ends of the interferometer, the phase accumulation is completely determined by the charge on the interferometer.

In Section IV, we discuss implications of our results for experimental measurements, and we briefly discuss  complications that arise when one tries to extend the results to quantium Hall states other than a Laughlin state or $\nu=1$.  Our results are summarized in Section V.}
{\color{black}  In the Appendix, we present supporting discussion of the effects of tunneling between the device edges and  localized states near the edge.}

\section{Effective edge theory}

\subsection{$\nu = 1$}

We begin by considering an infinite edge parallel to the $x$-axis, which separates  a $\nu = 1$ quantum Hall state in the region of positive $y$  from a vacuum state in the region of negative  $y$. We assume here that electron spins are completely aligned by the magnetic field, so we can ignore the electron spin for most of our discussion.

We shall be concerned here with the ground state of the system at a fixed value of the the magnetic field $B$ and a fixed value of the electrochemcal potential. We begin by working in a Hartree-Fock  approximation, where one is essentially considering a model of non-interacting electrons in a self-consistent potential, $V$, which depends on position.
Furthermore, we shall initially consider a translationally-invariant approximation, where we assume that $V$ is independent of $x$. 
We focus on a typical edge structure in a GaAs device. (Graphene devices may show significant differences in the microscopic edge structure \cite{graphene-edge}, 
though we still expect the same effective theory to describe the edge in the low-energy limit in GaAs and graphene.) 

In GaAs, the charge density changes gradually from zero in the depletion region to its maximal value $\nu_{\rm bulk}$ in the bulk. The width of the edge region, where the density changes, is much greater \cite{CSG,GH} than the magnetic length $l_B$. Within the translationally-invariant Hartree-Fock approximation, the gradual density change is accomplished by forming a series of alternating stripes\cite{recon}  of filling factors 0 and 1 with widths of the order of the magnetic length $l_B$, as illustrated in Fig. 2. The relative width of the stripes of the two filling factors, which  determines the average local charge density, will change over the width of the transition region.

Disorder modifies this picture in two ways. First, it creates a complex network of wiggling stripes of position-dependent widths \cite{D1,D2}. 
Second, the absence of translational symmetry in the $x$-direction destroys momentum conservation and allows charge tunneling between and across stripes. The second effect is illustrated in Fig. 2 by dashed lines.   {\color{black} In fact, external disorder may not be necessary here.  Within a Hartree-Fock approximation, one might be able to gain energy by breaking translational invariance in the $x$ direction and forming something like a Wigner crystal containing one electron per unit cell, which could open up an energy gap at the Fermi energy. 

In any case, we must take into account fluctuations beyond the Hartree-Fock approximation. The simplest  way to proceed is 
to consider each $\nu=1$ stripe in Fig. 2 as a one-channel quantum wire that can be understood using  a bosonized action \cite{Wen-book}}

\begin{eqnarray}
\label{1-phase}
L_n=\frac{1}{4\pi}\int dt dx [\partial_t\phi_{nL}\partial_x\phi_{nL}-\partial_t\phi_{nR}\partial_x\phi_{nR}+\mu_{nR}(x)\partial_x\phi_{nR}\nonumber\\
-v_{nL}(\partial_x\phi_{nL})^2-v_{nR}(\partial_x\phi_{nR})^2+\mu_{nL}(x)\partial_x\phi_{nL}],
\end{eqnarray}
where the index $n$ numbers the stripes (Fig. 2), $e\partial_x\phi_{nL}/2\pi$ and $e\partial_x\phi_{nR}/2\pi$ are the charge densities in the left- and right-moving chiral modes on the two sides of the stripe, $v_{nL,R}$ are the edge mode velocities, and $\mu_{nR,L}$ model the chemical potential and the coupling of the charge density with the random electrostatic potential. In general, there will be a finite number $N$ of stripes of finite width. The stripe with  $ n= N+1$ is the bulk $\nu=1$ state, which, here, stretches to $y= \infty$. Thus, in the region of interest, there will be $N+1$ right movers but only $N$ left movers. 

{\color{black}The action  (\ref{1-phase}) assumes a gauge of the form ${\bf{A}} = - B (0,x)$, so that the $A_x = 0$ for all the lines.} 

\begin{figure}[!htb]
\bigskip
\centering\scalebox{0.35}[0.35]
{\includegraphics{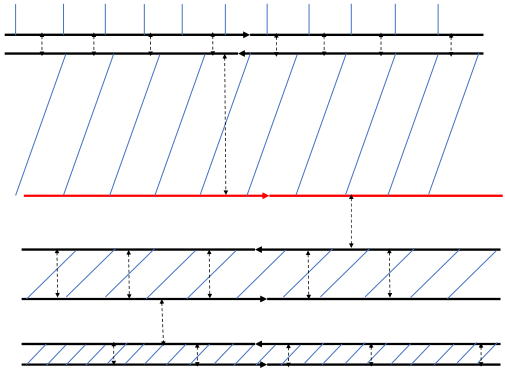}}
\caption{Shaded stripes are stripes of filling factor 1 separated by regions of zero filling factor. Each stripe carries two chiral edge channels on its two opposite edges. Dashed lines show electron tunneling. Tunneling across a stripe dominates in the lower part of the figure, where the average filling factor is low. Inter-stripe tunneling dominates at a high average density.
The red line is the delocalized chiral channel.}
\label{fig2:inter}
\end{figure}

We start with a model in which the only effect of disorder consists in random tunneling across the stripes and between neighboring stripes. The system is described by the action

\begin{equation}
\label{2-phase}
L=\sum_n (L_n -T_n-T_{n, n+1}),
\end{equation}
where $T_n$ describes tunneling across stripe $n$ and $T_{n,n+1}$ describes tunneling between stripes $n$ and $n+1$. The tunneling operators can be represented as

\begin{equation}
\label{3-phase}
T_n=\int dt dx u_n(x)\exp(i[\phi_{nR}(x)+\phi_{nL}(x)])+h.c.;
\end{equation}

\begin{equation}
\label{4-phase}
T_{n,n+1}=\int dt dx v_n(x)\exp(i[\phi_{n+1,R}(x)+\phi_{nL}(x)])+h.c.
\end{equation}
with random complex tunneling amplitudes $u_n(x)$ and $v_n(x)$. We assume that the mean squares of the amplitudes $\langle |u_n(x)|^2\rangle$ and $\langle |v_n(x)|^2\rangle$ do not depend on $x$ but depend on $n$. Indeed, a greater tunneling amplitude $u_n$ corresponds to a narrower $\nu=1$ stripe, while a greater tunneling amplitude $v_n(x)$ corresponds to a narrow interval between the stripes.  The spatially-varying complex phases of $u_n(x)$ and $v_n(x)$ are due to the magnetic flux enclosed between the edge modes.  {\color{black}In the gauge we have chosen, this results from the $x$-dependence of $A_y$.  In a  gauge where $A_ y=0$ and $A_x$ depends on $y$,  this would reflect the difference in the momenta of the electron states associated with these modes.}   
 
Consider first the limit in which $v_n$ and $v_{n-1}$ vanish for  stripe $n$ {\color{black} and let us ignore Coulomb interactions between neighboring stripes. Then the action for the chiral modes bounding the stripe is similar to that for the case of a non-chiral Luttinger liquid, and we expect that in the presence of  disordered
 tunneling across the stripe, the eigenstates will become localized. } Similarly, if $u_n$ and $u_{n+1}$ vanish, then the corresponding modes will be localized by inter-stripe tunneling $v_n$. We also expect  a tendency towards localization in the case where both $u_n$ and $v_n$ are present.  However, we know that at least one right-moving edge mode  must remain unlocalized, as  the non-zero edge conductance  of a quantum Hall state is topologically protected.  
 
 In the simplest case, the delocalized edge channel will coincide in space with one of the right-moving modes of the original translationally invariant model, as indicated by the red-colored mode in Fig. 2.  More generally, however, the weight of the delocalized mode may spread out, covering several of the original edge states, and the $y$-coordinate of its center may fluctuate significantly as a function of $x$.  
 
Within a Hartree-Fock picture, any given  localized state will have a discrete energy. The state will be occupied, if and only if its energy is below the Fermi level  $E_F$.  An empty state localized at a position with $y$ greater than that  of the delocalized edge state may be interpreted as a hole in the $\nu=1$ state, whereas an empty state with $y$ below that of the delocalized edge may be regarded simply as part of the vacuum.  Similarly, an occupied state with  $y$  well below the propagating state may be interpreted as an electron localized in the $\nu=0$ region, whereas a filled state with  $y$ above the edge mode may be regarded as simply part of the electron gas contributing to the filled Landau level. However, localized states that overlap in space with the extended state will have a more ambiguous interpretation.
 
{\color {black} In general, the requirement to minimize the total energy  will lead to changes in the self-consistent Hartree-Fock potential that tend to decrease the energy of occupied states and increase the energy of empty states.  This will tend to increase  energy gaps across  the Fermi energy, and will tend to increase localization of states other than the propagating extended state.  On the other hand, repulsive electron-electron interactions lead to screening of the disorder  potential, which can decrease the tendency to localization.}

 In  the bosonized description, one can envision eliminating the localized states by a renormalization group (RG) procedure as one moves towards lower energies.  Localized states appear here as closed loops which disappear as the running energy cutoff $E$  is reduced below the energy of the state. Typically, this will occur when the size of the state is of order $hv/E$, where $v$ is the edge velocity.   {\color{black} In some cases, however, a closed loop carries one state of energy $\ll hv/l$, which will not be eliminated by RG at the energy scale
 $hv/l$.}
 
 {\color{black} The situation at a late stage of the RG procedure is illustrated schematically in Fig. 3.  Three localized states are shown that are not yet eliminated and are close enough to the propagating edge that tunneling between them and the edge will affect the velocity of propagation along the edge after further renormalization.   
 
 The  effect on the propagating mode produced by a  localized state near to the edge can be understood using a simple model,  
 Consider a localized state whose energy $\epsilon_R$ is initially above $E_F$ by an amount which is large compared to its tunnel coupling to the edge, so that the state is empty. Then, imagine that the parameters are modified so that $\epsilon_R$ drops below $E_F$ by a large-enough amount that the state becomes occupied. If the Coulomb interaction is sufficiently screened,
 one can ignore additional effects produced by the change in the Hartree-Fock potential at the edge due to the change in occupancy of the localized state. One finds, however that during the process, the phase $\phi$ accumulated along the length of the renormalized edge state will have changed by precisely $2 \pi$.   
(An explicit solution of this model, in the case of non-interacting electrons, is reviewed in Appendix A, below.)
 This result applies regardless of whether the localized state was positioned in the $\nu=0$  region or the $\nu=1$ region, and it is a manifestation of  the condition stated in the definition of the action $L_n$,   that fluctuations in the one-dimensional charge density $\rho(x)$ associated with an edge are given by $e \partial_x \phi(x) / 2 \pi$.  
Solution of the model also shows that the way in which the phase shift varies for intermediate values of $\epsilon_R$ is given by arctangent curve with an energy spread of order
$\hbar \Gamma$, where $\Gamma$  is  the decay rate for an electron initially in the localized state to tunnel into the propagating mode, when $\epsilon_R$ is well above the Fermi energy.  
 
The cumulative effect of phase shifts due to interaction with multiple localized states can be characterized as a renormalization of the edge velocity by an amount which varies from point to point along the edge.  If one ignores all terms irrelevant in the renormalization group (RG)  sense, the action of the renormalized edge mode can then be written in the form 
 \begin{equation}
\label{L1}
L=-\frac{1}{4\pi}\int dt dx \partial_x\phi\{[\partial_t+v(x)\partial_x]\phi-\mu(x)\},
\end{equation}
where the edge velocity depends on the coordinate due to disorder effects. }

\begin{figure}[!htb]
\bigskip
\centering\scalebox{0.35}[0.35]
{\includegraphics{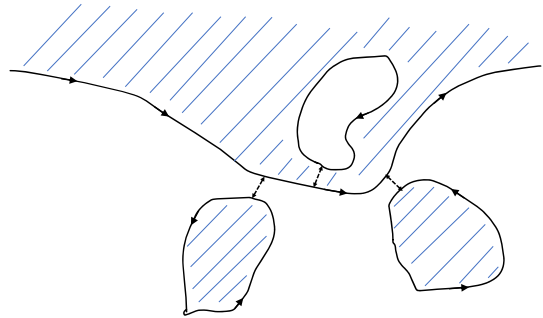}}
\caption{Sketch of a chiral edge channel  coupled  by tunneling to three closed edge channels.  Shaded regions are occupied by electrons at $\nu=1$. 
{\color{black}Arrows show the chirality of the edge modes.}} 
\label{fig3:inter}
\end{figure}

  \subsection{Laughlin states}

The arguments for $\nu=1$ can be extended to fractional quantized Hall states of the Laughlin form.  For example, consider a reconstructed $\nu=1/(2n+1)$ edge as a system of alternating $\nu=0$ and $\nu=1/(2n+1)$ stripes.  The low-energy theory after renormalization again reduces to the standard chiral Luttinger-liquid model \cite{Wen-book}.  The action is now 
\begin{equation}
\label{5-phase}
L=-\frac{1}{4\pi \nu   }\int dt dx \partial_x\phi\{[\partial_t+v(x)\partial_x]\phi-\mu(x)\},
\end{equation}
where the edge velocity depends on the coordinate due to disorder effects and we ignore all terms irrelevant in the RG sense.

A difference between a Laughlin state  and $\nu=1$ appears  if one considers the effect on the edge when the parameters of  a localized state in the interior of the interferometer are changed, so that a state initially above the Fermi energy now falls below the Fermi level.  In the case $\nu=1/3$, the change in the phase accumulation along the edge will be equal to $2 \pi / 3$.  However, the change $\delta Q $ in the electric charge when the localized state is filled  is that of a $\nu=1/3$ quasiparticle, namely $e/3$.  Therefore, in both the $\nu=1/3$ case and $\nu=1$ case the change in accumulated phase may be written 
\begin{equation}
\label{charge-phase}
e \delta [\phi(x_2)-\phi(x_1)]=2\pi  \delta Q,
\end{equation}

This result will be useful when we consider interferometry. {\color{black} We provide a detailed justification of this result in the next section.}

\section{Interference phase}

We now turn to a discussion of the Fabry-P{\'e}rot quantum Hall interferometer.  We consider only  the simplest case, sketched in Fig 1, where the bulk of the interferometer is in a simple quantum Hall state such as $\nu = 1$ or $\nu=1/3$, for which the ideal edge contains  just a single propagating mode.

We shall consider the possibility that the  edges are complicated by disorder and interactions, and that they may locally contain multiple forward and backwards propagating edge modes, but 
we shall assume that the constrictions are narrow enough that only one propagating mode enters and leaves from each side. 
We assume that the temperature $T$ and the voltage $V_S$ applied to source S are very low, and that as a result we can assume that particles passing through interferometer will not suffer inelastic processes. {\color{black} (Inelastic processes will be discussed in Subsection IV.A.)}      We shall also work in a limit where the tunneling probability at each constriction is small.

 We wish to calculate the total current $I$ that will be scattered from the lower to  upper edge of the interferometer and thus measured at contact  D, at a given small value of $V_S$.   We expect that $I$ will be proportional to $|t_1 +  t_2 e^{i \theta} |^2$, where $t_1$ and $t_2$ are the intrinsic reflection amplitudes for the two constrictions, and $\theta$ is the phase that would be accumulated by a  quasiparticle traveling in a closed loop  around interferometer region between the two constrictions. The accumulated phase will be sensitive to variations in parameters such as the applied magnetic field or to voltages $V_G$ applied to gates near the interferometer, so  if the parameters are varied, we expect to see oscillations in $I$ of the form
\begin{equation}
\delta I \propto   {\rm{Re}} (t_1^* \, t_2 \, e^{i \theta} ) .
\end{equation}
We shall argue that this phase accumulation can be related, after an appropriate gauge transformation, to the sum of ground-state  phase accumulations along the upper and lower edge. 

More importantly, however, we shall argue that the interference phase depends only on the total electron charge $Q$ within the interferometer region. To be more precise, let us draw 
 a closed contour $C$ which passes through the two constrictions and otherwise lies entirely in the vacuum outside the area of the interferometer, as represented by the dotted curve in Fig.~1. Then $Q$ may be defined as the total charge inside the contour $C$.  We  note that  an interferometer experiment only determines the phase $\theta$ modulo $2 \pi$, However, we argue below that for both $\nu=1$ and for a Laughlin state at $\nu=1/(2n+1)$,    it  is possible to define an absolute phase which will be  consistent with the interferometer measurements and is related to $Q$ by \begin{equation}
\label{Q-theta}
\theta =  2 \pi Q/e + \theta_0 ,
\end{equation} 
 where $\theta_0$ is a small phase shift that depends only weakly on parameters such as $B$ and $V_G$ and will therefore have no  measurable effect in an interferometer experiment.
 (We assume that the area of the interferometer is such that the number of contained flux quanta $N_{\Phi}$ is large, and we shall be concerned with oscillations in the electrical resistance caused by changes in the magnetic field or gate voltages which are large enough to change $N_{\Phi}$ by amount large compared to unity but small compared to $N_{\Phi}$. Thus, the characteristics of the constrictions can be assumed to be constant over this range.)

This result is compatible with the observation (\ref{charge-phase}) relating a change in the phase accumulation to a change in the electric charge associated with the edge.  However, it is more general, since it also applies to the charge jump produced by a change in the occupation of a localized state near the center of the interferometer that is not readily associated with either edge. {\color {black}  (Although the argument given in  Subsection III.A requires one to associate every localized state  with just one of the two edges, this is not required in the  more general argument of Subsection III.B.)}

 In the case of $\nu=1/3$, the addition of a quasiparticle to a localized state produces a charge jump of $e/3$, assuming that Coulomb effects on the edge can be ignored, which results  in a phase jump $\delta \theta = 2 \pi / 3$   that is observable in the interference pattern. 
This extra phase is just what one would expect from the fractional statistics of the quasiparticle. 

In a case where the Coulomb interaction between the localized charge and the edge mode is not well screened,  a change in occupation of the localized state may induce a change in the charge along the edge of the interferometer drawn from the reservoir beyond the contacts. The additional charge along the edge will be accompanied by an additional change in phase, so the total jump $\delta \theta$ will deviate from $2 \pi e^*/e$ in this case, {\color{black} where $e^*$ is one quasiparticle charge}.

 The power of the result (\ref{Q-theta}) is that it reduces the calculation of the interferometer phase, in the limit of $T \to 0$, to an equilibrium  calculation of the total electron charge in the interferometer. It is not necessary to ask whether the charge is located inside or outside the propagating edge state.  
 
 {\color{black}
 The result (\ref{Q-theta}) is similar to results appearing in the literature in other contexts.  (See,  for example, the work \cite{RoGe} of Rosenow and Gefen who consider effects on a Mach-Zehnder interferometer produced by interactions with a nearby quantum dot.)
{\color{black} A proof of (\ref{Q-theta}) for non-interacting electrons at $\nu=1$ follows standard ideas of scattering theory \cite{LL}. }  We shall give in Subsection III.B an alternate derivation, which is also
 applicable to the fractional case.   First however, we shall present an analysis of the interferometer experiment using the bosonization approach.}

\subsection{Bosonized Description} 
Within the bosonization formalism, we arrive at the following
model of the interferometer \cite{interferometry}:

\begin{eqnarray}
\label{7-phase}
L=L_{\rm u}+L_{\rm d}
-\int dt [\Gamma_1\exp(i[\phi_{\rm d}(x_1)+\phi_{\rm u}(x_1)])\nonumber\\
+\Gamma_2\exp(i[\phi_{\rm d}(x_2)+\phi_{\rm u}(x_2)])+{\rm h.~c.}]
\end{eqnarray}
at $\nu=1/(2n+1)$, where $L_{\rm u,d}$ are the actions of the upper and lower chiral edges with the charge densities $e\partial_x\phi_{\rm u,d}/2\pi$ and $\Gamma_{1,2}$ are complex amplitudes with the phases $\alpha_{1,2}$. 
{\color{black} The two exponents describe quasiparticle tunneling between the two edges of the interferometer.} We assume that $\Gamma_{1,2}$ are sufficiently small to allow the use of perturbation theory in inter-edge tunneling. 

{\color{black}Because we are now considering the action for a closed circuit around the interferometer region, the integral around the edge of the parallel component of the vector potential must  equal the  magnetic flux through the interferometer, so we can no longer assume a gauge where this component vanishes, as we did in the previous section. Then, in evaluating $L_u$ and $L_d$, the quantity  $\partial_x \phi $  in formulas such as  (\ref{1-phase}), 
 (\ref{L1}), and  (\ref{5-phase}), should be replaced by $(\partial_x \phi - {\color{black}\nu}e A_x )$  where the coordinate $x$ is taken to run along the edge.  After minimizing the action with respect to $\phi(x)$, this will result in a shift in the accumulated phase 
by the amount $e{\color{black}\nu} \oint A_x dx = {\color{black}\nu}e B A_c$ where $A_c$ is the area enclosed by the circuit.  Now the total charge in the interferometer will be the sum of the underlying bulk charge, $\nu e B A_c / 2 \pi$, and an additional edge  charge, given by  $e \oint  [\partial_x  \phi - {\color{black}\nu}e A_x ] dx  / 2 \pi$.  The contribution from $A_x$ drops out of the sum, so the total charge will be given by $e \Delta \phi / 2 \pi$, where $\Delta \phi$ is the total phase accumulated on the two edges of the interferometer.  

We  still have to relate $\Delta \phi$ to the phase measured in an interferometer experiment. 

The tunneling current between the upper and lower edges $I=dQ_{\rm u}/dt=i[T,Q_{\rm u}]/\hbar$, where $Q_{\rm u}$ is the charge of the upper edge and $T$ is the tunneling term in the square brackets in Eq. (\ref{7-phase}). Thus, the current operator is
\begin{eqnarray}
\label{8-phase}
I=\frac{i\nu e}{\hbar}[\Gamma_1\exp(i[\phi_{\rm d}(x_1)+\phi_{\rm u}(x_1)])\nonumber\\
+\Gamma_2\exp(i[\phi_{\rm d}(x_2)+\phi_{\rm u}(x_2)])-{\rm h.~c.}].
\end{eqnarray}
We note that after neglecting terms which are irrelevant at low energies, 
 $L_{\rm u,d}$ are quadratic with linear contributions in charge density that describe the effects of the chemical potential, the random potential, and the vector-potential component $A_x$. }
The average current, computed in the lowest order perturbation theory in $\Gamma_{1,2}$, is
\begin{eqnarray}
\label{9-phase}
\langle I\rangle=\frac{i}{\hbar}\int_{-\infty}^0 dt\langle [T(t),I(0)]\rangle=-\frac{\nu e}{\hbar^2}\int_{-\infty}^0 dt
\sum_{ij}\Gamma_i\Gamma_j^*\times\nonumber\\
\langle [\exp(-i\{\phi_{\rm d}(x_j,t)+\phi_{\rm u}(x_j,t)\}), \nonumber\\
\exp(i\{\phi_{\rm d}(x_i,0)
+\phi_{\rm u}(x_i,0)\})]\nonumber\\
-[\exp(i\{\phi_{\rm d}(x_i,t)+\phi_{\rm u}(x_i,t)\}), \nonumber\\
\exp(-i\{\phi_{\rm d}(x_j,0)+\phi_{\rm u}(x_j,0)\})] \rangle,
\end{eqnarray}
where the angular brackets denote the average with respect to the quadratic part of the action, {\color{black} $L_u + L_d$, under conditions where there is a small voltage difference  $V_S$ between the lower and upper edges of the interferometer.  The voltage difference may be implemented by shifting the chemical potential $\mu(x)$ on the lower edge, relative to its equilibrium value,  by the amount $-eV_S$.

It is convenient to write $\phi(x,t) = {\tilde{\phi}} (x,t) + \eta(x, t)$,  where $\eta(x,t)$ is the mean value of $\phi(x,t)$ in the absence of tunneling. The quantity $\eta(x,t)$ will actually be independent of $t$ except for a term which is constant along each edge but which increases linearly in $t$ with  a rate  that depends on the voltage of the edge.  Then ${\tilde {\phi}}(x,t) $ is a Gaussian variable with zero mean.  If the current $\langle I \rangle$ is expressed  in terms of ${\tilde{\phi}}$, it will have the same form as  (\ref{9-phase}) except that 
the phases of $\Gamma_j$ will be shifted by an amount which multiplies the terms proportional to  $\Gamma_1 \Gamma_2 ^*$  and  $\Gamma_1 ^* \Gamma_2 $ by a factor $e^{\pm i \theta_a } $, where  
\begin{equation}
\label{theta-phi}
{\color{black} \theta_a=\int dx \, \partial_x(\eta_{\rm d} + \eta_{\rm u})=\int dx \, \langle \partial_x(\phi_{\rm u}+\phi_{\rm d})\rangle .}
\end{equation} 
Fluctuations in ${\tilde{\phi}}$ will renormalize the magnitudes of the various terms in the current, but they will not affect  their phases. 

From our previous arguments, we expect that the right-hand side of  (\ref{theta-phi}) should be equal to $2 \pi Q / e$, where $Q$ is the expectation value of the charge on the interferometer in its ground state.  Thus the measured interference phase $\theta$ should have the form  (\ref{Q-theta}), where $\theta_0$ is the phase of the original product $\Gamma_1 \Gamma_2 ^*$

Little changes when we go beyond the lowest order of the perturbation theory. Indeed, the terms of the order $|\Gamma_1|^{2n}|\Gamma_2|^{2m}\Gamma_2^p\Gamma_1^{*p}$ are proportional to $\exp(ip\theta)$. }

\subsection{Relation between phase and charge}

Suppose that we modify the geometry shown in Fig.~1 by removing the contacts, so the device is now a closed system with conserved electron charge.  We shall assume that regions to the left of constriction 1 and to the right of constriction 2 are ideal quantized Hall systems, with clean edges and no impurities, with filling factor $\nu=1$ or a Laughlin state with $\nu=1/(2n+1)$. We shall further assume that the areas $A_L$ and $A_R$ of the outer regions are  finite but very much larger than the area of the interferometer region itself.  We focus our attention on the  ground state of the system with a total electron number $N$, which will be a large integer.   If we denote the  expectation values of the electron number in the outer regions as $N_L$ and $N_R$, and the number inside the interferometer region by $N_{\rm{in}}$, we know that their sum must be equal to the integer $N$, but the individual numbers are not quantized, as electrons can flow between the three regions.  

Since the outer regions are ideal, their low-energy degrees of freedom are associated with a single edge state,  and they will be well described by a bosonized edge mode, with a quadratic Lagrangian of the form assumed in Section II. The interferometer region may be far from ideal, but if only a single edge mode enters the constriction at each side, the phase variables $e^{i \phi_u(x)}$ and $e^{i\phi_d(x)} $ should be well defined at the contacts. Furthermore these phases must be continuous  across each constriction.  If we define a quantity 
\begin{equation}
{\color{black} \theta = \phi_u(x_1) -\phi_u(x_2) - \phi_d(x_2) + \phi_d(x_1) },
\end{equation}
the requirement that the phase factors are continuous and single valued means that $\theta$  plus the phase accumulations along the exterior of $A_L$ and $A_R$ must add up to zero, modulo $2 \pi$.  

Since the outer regions are ideal, we know that the phase accumulations around the regions $A_L$ and $A_R$ will be equal to $2 \pi$ times $N_L$ and $N_R$, respectively.  The edges of the interferometer region may be far from ideal, {\color{black} but in  any case, we would expect that  if $B$ and all gate voltages are fixed, the values of $\theta$ and  $N_{\rm{in}}$ will be determined by the electrochemical potential of the system. }

Now let us apply to the system a weak external potential  $V(\mathbf{r} )$,  which is chosen  to be zero inside the interferometer region but different from zero in at least some part of the outer regions, including a portion of their edges.  If $N$ is held fixed, $V$ will generally change $N_L$ and $N_R$ and it  will change the electrochemical potential of the system by an amount such that $\Delta  N_{\rm{in}} + \Delta  N_L + \Delta N_R = 0   $.  Similarly, the change in $\theta$ mod $2 \pi$ will be equal and opposite to the change in the phase accumulation in the ideal  outer regions.  This implies that as the electrochemical potential is changed,  the relation $\Delta \theta = 2 \pi  N_{\rm{in}}$ mod $2 \pi$ will be maintained. The relation will similarly be maintained if we vary the magnetic field or the gate voltages on the interferometer, as long as the constrictions and the outer regions remain in the quantized Hall state $\nu$ and as long as the interferometer remains in a state where we can neglect scattering between the upper and lower edges. 

With this reasoning, we have established that the ground state phase accumulation $\theta$ will satisfy Eq. (\ref{Q-theta}) modulo $2 \pi$.  One must still show that the phase measured in an interference experiment at low energies is the same as the ground state  phase. {\color{black}Here we need to make the additional assumption that 
 the rate for scattering of a  charged quasiparticle between the upper and lower edges is much smaller than the inverse   of the propagation time along an edge.  Therefore, current is conserved separately on each edge, and the phase accumulation on each edge will be fixed by the electrochemical potential on that edge.  If $V_S$ is sufficiently small, the phase accumulation will be the same as in the ground state.

We now follow} the same reasoning as was used in subsection III.A. Although the phase variable $\phi (x)$ may not be well defined at all points along the edge of the interferometer, it should still be valid to assume that quantum fluctuations in the phase difference between the two ends of the edge are small, after high frequency fluctuations are eliminated, and that the remaining fluctuations should be approximately Gaussian.

Since the interferometer phase and the ground state phase $\theta$ are, physically,  only defined modulo $2\pi$, we are free to choose $\theta$ to satisfy (\ref {Q-theta}) absolutely.

\section{Implications for Experiments at $\nu=1$ and $1/3$}

Interferometry experiments have typically been interpreted in terms of a simplified model that assumes the existence of one or more propagating edge modes near the interferometer boundary, together with a set of localized states that have  only negligible tunnel coupling to the propagating modes.  In the case of $\nu=1$ or $1/3$, there is only one propagating mode at each edge.  Since the propagating modes in the interferometer region pass almost freely through the two bounding constrictions, their electrochemical potential is set by the voltage on the external leads. Consequently, their associated charge densities can vary continuously as a function of the magnetic field and the electrostatic potentials arising from voltages on gates or from Coulomb interactions with localized charges.  

In the case where the Coulomb interactions are well screened, changes in the occupation of localized states do not  affect  the charge densities on the propagating edge states.  Then, for $\nu=1$,  where  filling of an occupied state causes $\theta$ to jump by $2\pi$, such a jump has no effect on the interference signal.  In the case of $\nu=1/3$, localized states inside the propagating edge mode can be occupied by quasiparticles with charge $e/3$, so the occupation of such a level will produce an observable jump in $\theta$ equal to $2 \pi /3$.  By contrast, localized states outside the edge mode can only be occupied by electrons with charge $e$, so they have no measurable effect, and they can be ignored, just as for $\nu=1$.  

In reality, however, we expect to find a certain number of localized states that are so close to the propagating mode  that tunneling to them cannot be ignored. 
 As discussed in Appendix A,   for a single localized state near a $\nu=1$ edge, tunneling will lead to an energy broadening of the state by an amount proportional to  the decay rate $\Gamma$  for  charge on the localized state to equilibrate with the propagating edge. The occupation of the localized state at $T=0$ is then no longer restricted to be 0 or 1, but can vary continuously as the parameters are varied. Similarly, the associated change in the accumulated phase $\theta$ will not be discontinuous but will be spread out by an amount proportional to $\Gamma$. 

In the same way, at $\nu=1/3$, for a localized state that overlaps strongly the propagating edge mode, the average charge in the ground state could vary continuously with the parameters of the system. We may think of the ground state as a linear  combination of two quantum states whose instantaneous occupations differ by one. The relative amplitudes of the two states will change as the parameters are changed. In addition, for  the case of $\nu=1/3$, as the center position  of localized state  is moved outside the position of the propagating mode, the accumulated  electric charge difference associated with a unit change in occupancy will  also vary 
from $e/3$ to $e$.  (In the composite-fermion-Chern-Simons picture \cite{Jain-Book}, this may be understood from the result that the net change in charge density produced by the Chern-Simons flux attached to a composite fermion depends on the background charge density in the vicinity of the particle.) 

In the experiment of Ref.~\onlinecite{manfra20},   the  sample parameters were initially tuned to a region where the bulk of the interferometer was in an incompressible state at $\nu=1/3$.  Specifically, the Fermi level was inside the energy gap of the ideal $\nu=1/3$ state, and the density of localized states in the gap due to impurities in the bulk  was  small. Consequently, parameters such as the magnetic field or the voltage on a side gate could be varied over a range large enough to observe several oscillations in the interferometer signal before there was a change in  occupation of a localized state far from the boundaries.  
Over a more extended range of parameters, however, several jumps in phase by amounts close to $2 \pi /3$ were observed, which were interpreted as arising from changes in occupancy of a localized state  far from the edge as its energy passed through the Fermi level.  {\color{black} This is in accord with the predicted effect due to the fractional statistics of an $e/3$ quasiparticle  in a simple model with an ideal edge state. However, we would like to understand whether this might be altered in a more realistic model.

In the region near the edge where the electron density varies  from the bulk density to zero, we expect the density of localized states to be large, of the order of one state for every few flux quanta. } {\color{black} Of course, not all these states are low enough in energy to matter for our discussion.} But if the positions of these localized states and their energies were random and  fixed by external disorder, one might expect to see large fluctuations in the accumulated charge, and consequently the interferometer phase, as the position of the propagating mode is varied due to changes in the gate voltage or magnetic field.   However, we actually expect that the density of localized states, and even their locations, will be dominantly set by the self-consistent potential arising from electron-electron interactions, and these states will move in or out along with the propagating mode as external parameters are varied.  Then in the absence of external disorder, the total charge on the interferometer would vary smoothly, effectively linearly, with variations in $B$ or $V_G$, giving rise to a simple periodic variation of $e^{i \theta}$. 

 {\color{black} The effects of external disorder would lead to some residual fluctuations in the phase. We do not expect these effects to be large, due to electrostatic screening and the long-range nature of the dominant disorder. Even more importantly, however, as we show below, 
the interferometry phase is insensitive to local details such as disorder as long as screening is linear and can be described in terms of a capacitance.}
{\color{black} Indeed,} because the distance $d$ to the screening gates is large compared to the magnetic length $l_B$,  we  expect that  the quantities $\partial Q/ \partial B$ and $\partial Q/ \partial V_G$, which  describe the response of the electric  charge, and hence of $\theta$,  to changes in the magnetic field or side-gate voltage, will be dominated by geometric electrostatic effects, and will be relatively insensitive to details of the edge.  Similar considerations apply if one wishes to compute the change in $Q$ that would result from a change in the {\color{black} electrochemical potential  set by the leads, while $B$ and $V_G$ are held fixed.  

Suppose that we apply a small voltage $V$ to all of the leads, and} let us define an effective  edge capacitance by  $C_e = \partial Q / \partial V$ with  bulk charge held fixed.  Then we may write
\begin{equation}
\frac {1}{C_e} = \frac {1}{C_G}  + \frac {1} {C_Q}\, , \,\,\,\,\, C_Q \equiv   e \frac  {\partial Q} {\partial \mu} ,
\end {equation} 
{\color {black} where $C_G$ is the geometric capacitance that would be present if the compressible edge region were treated as a perfect conductor, and the quantum capacitance $C_Q$ is the correction due to the finite compressibility.  Specifically,  $\partial Q / \partial \mu$ measures}  the change in edge charge density that would be produced by a change in chemical potential if the long-range part of the Coulomb interaction were completely screened.  For $d/ l_B \gg 1$, we expect that  $C_G$ will be smaller than $C_Q$, so that $C_e$  will be only slightly affected if we treat the compressible region as a metal, taking  $C_Q \to \infty$. 

The edge capacitance $C_e$ is also  important for understanding the decrease in the interference signal when  the temperature is raised above zero.  Thermal fluctuations will lead to a Gaussian distribution of the edge charge, with variance  given by 
\begin{equation}
\langle (\delta Q)^2 \rangle =  C_e T .
\end{equation} 
{\color{black} Then the thermal expectation value of the quantity
 $\exp(i\theta)$, which enters the interference signal, will be reduced by a factor}
\begin{equation}
\langle e^{2 \pi i  \delta Q/{\color{black}e}} \rangle =  \exp ( - 2 \pi^2 C_e T / e^2).
\end{equation}

We  note that the velocity $v$ for propagation of a long-wavelength charge fluctuation along the sample edge is  determined by the Hall conductance and the capacitance per unit length.
{\color{black} Specifically, the action (\ref{5-phase}) implies the energy
\begin{equation}
\label{energy-phase}
E=\frac{\hbar}{4\pi\nu}\int dx v(x) \left[\frac{2\pi\rho(x)}{e}\right]^2,
\end{equation}
where $\rho(x)$ is the charge density. The energy minimum at a given charge $q=\int dx\rho(x)$ is achieved for $\rho(x)=q/[v(x)\int {dx'}{v^{-1}(x')}]$.}
 As a result, one finds that
\begin{equation}
C_e = \nu e^2 L h^{-1} \langle v^{-1} \rangle_e ,
\end{equation}   
where $L$ is the perimeter of the interferometer region and $ \langle ... \rangle_e $ denotes an average along the edge.

The experimental results of  Ref.~\onlinecite{manfra20}, in the region where the sample bulk remained in an incompressible $\nu=1/3$ state, were consistent with the arguments  presented  above. Specifically, a color plot of the interference signal as a function of  $B$ and $V_G$ showed a series of  stripes of negative slope.  The distance between successive conductance maxima  in any direction in the plane may be understood as the change in $B$ and/or $V_G$  necessary to add one electron to the interferometer.  The experimental result for the spacing in $B$, with $V_G$ held fixed, was found to be
\begin{equation}
\Delta B = 3 \Phi_0 / A_0 ,
\end{equation} 
where $\Phi_0 = h/e$ is the magnetic flux quantum, and $A_0$ is, at least approximately,  the area of the interferometer.  This is just the amount necessary to add one electron to the system if the area remains fixed and the bulk electron density is pinned to the ideal value, $\nu B / \Phi_0$.   The observed pattern of parallel stripes was interrupted by a series of phase jumps of size $2 \pi /3$, occurring along certain lines of positive slope, which are understood as arising from changes in the occupation of  a small number of localized states due to disorder deep in the interior.  {\color {black} According to our analysis, one should identify $A_0$ with $\nu^{-1} \Phi_0 \partial Q/ \partial B $, under conditions where there are no changes in the charges on localized states in the interior incompressible region. This should not be very different from the value   $\nu^{-1} \Phi_0 Q/B$. }

The resistance measurements of   Ref.~\onlinecite{manfra20} extended beyond the region where the Fermi level was inside the bulk energy gap of $\nu=1/3$ into the compressible
regions on either side.  There, one is in a region where there is a large density of  either positive or negative localized quasiparticles and therefore of  localized states at the Fermi level. However,   the longitudinal conductivity $\sigma_{xx}$ in the bulk remains small, and the bulk Hall conductance remains at the quantized value,  because the localized quasiparticles give a vanishing contribution  to transport.  In these  compressible regimes, the interferometer oscillations with $V_G$ remained visible, but oscillations with $B$ were much weakened and only visible at the lowest temperatures. This behavior can be understood in the simple model,  \cite{review-FH,ros-ste}  if   one takes into account the systematic variation of the occupation of localized states in the bulk with changes in $B$ or $V_G$.

As was noted above, for $\nu=1$, a change in the occupation of  a localized state in the bulk will have no observable effect on the interferometer signal if the Coulomb interaction between charges in the bulk and the edge is well screened.  However in samples without a nearby gate, Coulomb interactions can be important.  Although the Coulomb effect of  adding a charge to the bulk of a truly insulating system would be expected to depend importantly on  the location of  charge, particularly on the distance from it to the nearest edge, the situation can be quite different in a case where there is a large density of localized states at the Fermi energy.  Even if $\sigma_{xx}$ is negligibly small compared to the quantum conductance $e^2/h$, it may be big enough so that equilibration of the charge density can occur on the time scale over which parameters such as $V_G$ or $B$ can be varied.  Then the bulk of the  two-dimensional electron system will behave like a floating metallic layer \cite{CSG}, which will screen the localized charge, reaching a new uniform  potential determined by its overall capacitance to ground. 

Within this model, it was argued that if coupling between the bulk and edge is relatively weak (the ``Aharonov-Bohm'' regime) the dominant interference period should be a magnetic-field dependence corresponding to the addition of one flux quantum\cite{review-FH,int2,ros-ste} to the area $A_0$.  However, in the opposite (``Coulomb-dominated'') regime, where the bulk-edge coupling is strong, the most important oscillations in the interference pattern should have a period in $V_G$ corresponding to the addition of one electron to the system, but should be essentially independent of $B$.  For intermediate values of the bulk-edge coupling, the two interference patterns may be simultaneously visible\cite{int2}.

{\color{black}
\subsection {Inelastic processes.}  
In our discussion of the interferometer process, we have, so far, neglected the possibility of inelastic scattering.  
In the case of electrons at $\nu=1$, inelastic process occur if an electron 
tunneling across the constriction does not result in  a single electron with the same energy  traveling back along the second edge, but instead results in several particles and hole excitations, along one or both edges.  For the Laughlin states, where quasiparticles along the edge are strongly interacting, it is better to define inelastic processes as occurring when the final state contains additional excitations of plasmons along one or both edges. Within the bosonization formalism, such terms can arise from terms beyond the quadratic in the effective Lagrangian.  

According to the RG analysis, in the absence of tunneling,  at  $T=0$, the quadratic Lagrangian should be adequate to describe propagation along the two edges 
of the interferometer. Even if there is a large voltage difference $V_S$ between the edges, each edge will be in the ground state  appropriate to its own electrochemical potential. However, once tunneling occurs, there will be out-of-equilibrium quasiparticles moving, along the edges, and these could possibly create additional excitations.  At finite temperatures, even in the absence of tunneling, there could be inelastic processes along the edges, if the quadratic approximation is no longer valid. Of  course, if $V_S$ is made large compared to the energy scale $ \hbar v /L $ the interference signal will be lost, even in the absence of inelastic scattering, as particles of different energies will undergo different phase accumulations as they
 move around the interferometer. \cite{interferometry}

The existence of a large density of localized states near the edge, which might occur in the case of strong disorder and soft confinement,  could possibly lead to additional decoherence effects due to inelastic scattering processes, for measurements at non-zero bias voltages or temperatures.  Low-energy neutral excitations could arise from processes where a quasiparticle is taken from an occupied state below the Fermi energy and transferred to a nearby empty state, just above the Fermi energy.  More complicated excitations may involve multiple exchanges. Eventually, these excitations would reradiate their energy into plasmons or particle-hole excitations along the edges. The matrix elements by which charge propagating along the edge can exchange energy with the localized modes are likely to be small at low energies, but their effects on decoherence deserve further study.

}

\subsection {Other filling factors}

The arguments for the robustness of the interference phase given above do not apply directly when the bulk of the interferometer belongs to a Hall state other than $\nu=1$ or a Laughlin state with $\nu = 1/(2n+1)$,  so  the interferometer edge will contain more than one propagating mode.  In this case, the interferometer pattern will depend importantly on which mode  is partially reflected  at the constrictions.  

The simplest case to consider is  when the  bulk has nominal filling $\nu=2$.  Then the ideal edge will have two co-propagating modes: an outer edge mode corresponding to electrons with the majority spin and an inner edge mode belonging to the minority spin.  If  the constrictions are not too narrow, both edge modes can be transmitted through them.  If we can neglect spin-orbit coupling then there will be separate phase accumulations $\theta_\sigma$ for the two spin states, which will be related to the total  charge $Q_\sigma$ in the interferometer region of electrons 
with the given spin.  Since we expect, in this case, that back-scattering at the constrictions will occur primarily for electrons in the inner channel, the interference signal should then be determined by the number of minority-spin electrons in the system.  Because electric charge can be redistributed between the two edge modes, the responses of the system to changes in $B$, $V_G$, or $V$ will not be determined only by relatively long-wavelength electrostatic energies, and they may be much more sensitive to edge disorder or other details than in the case of $\nu=1$.  Also, since one expects to find a spin-wave edge mode with velocity much smaller than the charge velocity, the interference signal may disappear more quickly with rising temperature than for $\nu=1$. 

{\color {black} Other issues occur if the bulk filling is on a $\nu = 2$ Hall plateau, but the constrictions are so narrow that the minority carrier mode is completely reflected, while the majority spin mode is partially transmitted. Because the number of minority-spin electrons is now constrained to be an integer, the interferometer phase can be determined equally well from the number of majority spin electrons or the total electron number. However, the existence of two spatially separated edge modes may make it more difficult to approximate  the edge region by a single geometric capacitance.  }

For integer fillings larger than $\nu = 2$, disorder at the edge can cause additional complications due to  scattering between edge modes with the same spin.  The possible
interferometer patterns for integer states with $\nu \geq 2$ have been discussed in the literature\cite{int1}, but primarily within a  a model that neglects  scattering between edge modes. 
A similar discussion, which ignores scattering  between edge modes, was applied to fractional quantum Hall states, such as $\nu=2/5$ or 3/7, with multiple edge modes all traveling in the same direction \cite{int2}. It is not clear whether predictions of these models will apply to actual physical systems where such scattering may be important.

\section{Summary}

Experiments with the geometry of a Fabry-P\'erot interferometer have been used to study various aspects of the quantum Hall effect, including the fractional charge and fractional statistics of quasiparticles. However, interpretation of these experiments have tended to use a somewhat oversimplified  picture, wherein there is a set of one or more propagating modes at the sample edge and a set of localized quasiparticle states that have only negligible tunneling connections to the propagating modes.  

In this work, we have explored the question of how robust are these predictions in a more realistic model, where the propagating edge mode is likely to be embedded in a region containing a large density of quasiparticle states, which will  be coupled to the edge with tunneling probabilities of various strengths.
Our focus was on the simplest quantized Hall states, $\nu=1$ and Laughlin states at $\nu= 1/(2m+1)$, which have a single chiral edge mode in idealized models.   

We first considered the edge of a  semi-infintite sample.  We started from  a microscopic picture of the edge environment in the presence of disorder, and we discussed how a single propagating mode emerges at low energies, both in a Hartree-Fock picture and with a more accurate analysis using bosonization of the edge modes. This led to a description in terms of a single renormalized edge mode, with a well defined phase accumulation as one moves along the edge.  Moreover,  changes in this accumulated phase could be related  to changes in the electric charge associated with the edge. 

In Section III, we turned to measurements in an interferometer geometry. We showed how the measured phase of oscillations in the interferometer signal  at low energies can be directly related to the accumulated phase change $\theta$ around the edge of the interferometer in its ground state,  and that $\theta$, in turn, is related  by Eq. (\ref{Q-theta}), in a precise way, to the total electric charge in the interferometer region.   This relation holds equally for the Laughlin states as for the integer state at $\nu=1$.  

In Subsection III.B,  we presented   {\color{black}  a more general  detailed argument   for the validity of Eq. (\ref{Q-theta}) in the ground state,
assuming  that the constrictions defining the interferometer area are smooth and allow just a single chiral mode to pass through, but making no other assumptions about the behavior of the edge states or other details of the system inside the interferometer.   We then  argued that the ground state value of $\theta$ would be manifest in  interference measurements at sufficiently low energies, provided that  the body of the interferometer is close enough to a quantized Hall plateau that $\sigma_{xx}$ is small and one can  neglect scattering  from one edge to the other across the bulk of the interferometer. }  

{\color {black} Some consequences of  this  analysis  were explored in Section IV. 
One consequence  is a confirmation}  that the addition or removal of a Laughlin quasiparticle from a localized state in the interior of  the interferometer region will produce precisely the change in interferometer phase as is predicted by the theory of fractional statistics, provided that  Coulomb interactions between the localized quasiparticle and the edge mode can be neglected because of  screening by a nearby conductor.  More generally, Coulomb interactions should be taken into account. However, the effects of changes in parameters such as the magnetic field or a gate voltage, as well as effects of a discrete change in occupancy of a localized state, will still be completely  determined by changes in the total charge on the interferometer. This charge  can be obtained, in principle, from a generalized electrostatics calculation, where one takes into account the finite  local compressibility of the electron system in quantized Hall regions with a varying density of localized quasiparticle states.   However, in many situations, the quantum energy cost of charge fluctuations in a compressible region is 
negligible  compared to the electrostatic energy cost, so that compressible regions can be treated as ideal conductors. {\color{black} We also argued that under conditions where the number of localized states is held fixed,  the variation in interference phase with changes in the applied magnetic field should be only weakly affected by disorder or other details of the edge: the variation should be smooth and  essentially determined by the area of the interferometer. }  

Quantum Hall states with multiple chiral edge modes are harder to deal with, particularly if one cannot neglect scattering between  the different modes.  Nevertheless, insights gained from our analysis may be helpful in understanding these systems. {\color{black} In particular, the expectations of the naive model should hold if the bulk-edge interaction is sufficiently well screened, and scattering between different edge modes can be neglected.}

\section*{Acknowledgments} DEF was supported in part by the NSF under grant No. DMR-1902356.  The authors have benefited from discussions on this subject with many colleagues over the years, and this work was motivated in part by questions they raised. 
Among these colleagues were  Moty Heiblum, Jainendra Jain, Philip Kim, Charles Marcus, Izhar Neder, Bernd Rosenow, Ady Stern, Tom Werkmeister, and Amir Yacoby.

\appendix

\section{Broadened phase jumps}

We  consider here the way the phase accumulation changes as one changes the relative energies of a localized state and the edge state, when the tunnel coupling between the two is not negligible.
In the simplest picture, small changes in the gate voltage and the magnetic field generally result in small changes in the device area. Hence, away from the $2\pi\nu$ jumps, the phase changes linearly with the voltage and the field. The existence of localized states in and out of the interferometer complicates this picture. Indeed, as we saw in Section II,
the low-energy effective theory contains tunneling between the chiral edge and localized low-energy states. The states outside the interferometer can exchange electrons with the edge. The states inside the interferometer can exchange quasiparticles with the edge. We will focus on $\nu=1$, where quasiparticles and electrons are the same. Our current goal is to see what happens to the phase when a localized state merges with the interferometer bulk in response to changing parameters. We will see that the phase exhibits a glitch of a Lorentzian shape. The total change of the phase in the process of merging is, of course, proportional the total charge in the localized state, which is constrained to be an integer number of quasiparticles.

We shall address the effect of a single impurity state that is close enough to the edge so that we need to take into account tunnel-coupling to the edge, {\color{black} but we shall neglect electron-electron interactions.}  We consider a closed edge with length $L$ and velocity $v$. We treat the impurity state as a resonant state with a bare energy $\epsilon_R$. The Hamiltonian is (with $\hbar=1$)

\begin{equation}
\label{11-phase}
H = H_0 + H_R + H_{\rm int}, 
\end{equation}
\begin{equation}
\label{12-phase}
H_0 =\sum_n v k_n a^\dagger_n a_n,
\end{equation}
\begin{equation}
\label{13-phase}
H_R = \epsilon_R c^\dagger c,
\end{equation}
\begin{equation}
\label{14-phase}
H_{\rm int} = \frac{1}{\sqrt{L}}\sum_n (\gamma a^\dagger_n c+\gamma{\color{black}^*} c^\dagger a_n),
\end{equation}
where $k_n=2\pi n/L$ with $n_{\rm min} \le n \le n_{\rm max}$. For simplicity we will choose symmetric cutoffs, $n_{\rm min}=-n_{\rm max}$, and will assume these to be so large that they can be sent to infinity.  This comes at the price of resolving the delta-function in the coordinate representation of 
$H_{\rm int}=\gamma \int dx a^\dagger(x) c\delta(x)+h.c.$, which we resolve as a function $u(x)$, nonzero at small $x$ only and with a unit integral.

For non-interacting electrons,  we need only  consider the behavior of a single electron in the system. We introduce the operator $\Psi$, which annihilates the eigenstate of $H$ with energy $E$,
\begin{equation}
\label{15a-phase}
\Psi =\int dx \psi^* (x)a(x) +\alpha c,
\end{equation}
where $\psi(x)$ is a wave-function.
Computing the commutator of $\Psi$  with the Hamiltonian, one gets
\begin{equation}
\label{16a-phase}
E\alpha=\epsilon_R\alpha+\gamma A,
\end{equation}
\begin{equation}
\label{17a-phase}
A=\int u(z)\psi^* (z)dz,
\end{equation}
\begin{equation}
\label{18a-phase}
E\psi (x)=-iv\partial_x\psi (x)+\alpha^*\gamma u(x).
\end{equation}
The solution for $\psi(x)$ {\color{black} can be obtained with the multiplication of Eq. (\ref{18a-phase}) by the integrating factor $\exp(-iEx/v)$. The solution is

\begin{equation}
\label{19a-phase}
\psi(x)=\exp(iqx)[C-\frac{i\Gamma A^*}{E-\epsilon_R}\int^x_{-\infty}\exp(-iqy)u(y)dy],
\end{equation}} 
where $C$ is a real normalization constant, $q=E/v$, and $\Gamma=|\gamma|^2/v$. Neglecting $qx$ in the region of nonzero $u(x)$,
{\color{black} we can connect $A$ and $C$ with the help of Eq. (\ref{17a-phase}).
The calculation involves the integral $I=\int_{-\infty}^{\infty} dz u(z) \int_{-\infty}^z u(x) dx=\int_{x<z}dx dz u(x) u(z).$
Since $I=\int_{z<x}dxdz u(x)u(z)$, it follows that $2I=\int_{-\infty}^\infty dz u(z)\int_{-\infty}^\infty dx u(x)=1$.
If one substitutes (\ref{19a-phase}) into (\ref{17a-phase}) and neglects $qx$ in the region of non-zero $u(x)$, one is led to the result $A[E-\epsilon_R – i \Gamma/2] = C(E-\epsilon_R)$. From this, one finds that  the phase shift  $\Delta \theta$ for an open system at energy $E$ is given by:

\begin{eqnarray}
\label{17-phase}
\exp({i  \Delta\theta} )=\frac{E-\epsilon_R-i\Gamma/2}{E-\epsilon_R+i\Gamma/2}, \nonumber\\
\Delta \theta = 2 \arctan \frac{\Gamma }{ 2 (\epsilon_R – E)}.
\end{eqnarray}

In the case where there are multiple impurities, one should add the phase shifts, or multiply the phase factors. The phase shift divided by $2\pi$ yields the excess number of states with the energies below $E$ compared to the absence of the impurities \cite{LL}.
The correction to the interferometry phase at a low temperature corresponds to substituting the chemical potential in place of $E$. For a single impurity, the phase changes by $2\pi$ as $E$ runs from a large negative value to a large positive value.

{\color{black} We expect that localized states are typically associated with extended closed edge channels with the length scale $\sim 100$ nm set by the distance to remote ionized donors. The relative energy of the interferometer and the localized state depends on the gate voltage in a way sensitive to the geometry of the device and the localized state.}

\end{document}